# Delay Adaptation Method for Relay Assisted Optical Wireless Systems

**Yu Zeng, Sanaa Hamid Mohamed, T. E. H. El-Gorashi and Jaafar M. H. Elmirghani**
*School of Electronic and Electrical Engineering, University of Leeds, LS2 9JT, United Kingdom*
*e-mail: ml16y5z@leeds.ac.uk, elshm@leeds.ac.uk,*
*t.e.h.elgorashi@leeds.ac.uk, j.m.h.elmirghani@leeds.ac.uk*

**ABSTRACT**
In this paper, we investigate optical wireless repeaters as relay terminals between a transmitter and a user in an Infrared Optical Wireless Communication (IROWC) system. A delay adaptation method is introduced to solve the problem of irregular signal arrival time from different relay terminals. Three different relay terminal deployment scenarios were investigated in a typical two-phase relay IROWC system with the proposed delay adaptation method. The simulation results indicate that the proposed system has better impulse response compared to the conventional system and that the root-mean-square delay spread of the relay system with the delay adaptation method is on average 30% less than the conventional system.

## 1. INTRODUCTION

Relay communication techniques can be attractive in Infrared Optical Wireless Communication (IROWC) systems. IROWC systems are a kind of indoor OWC systems that are not affected by dimming control and illumination requirement [1]-[8]. Such systems can be considered as complementary to conventional radio frequency communication (RFC) systems or as independent systems to provide wireless communication in indoor environment [9]-[15]. Conventional diffuse systems (CDS) are a form of IROWC system that contain a diffuse transmitter and wide field of view (FOV) receiver at the user device. Due to Inter Symbol Interference (ISI) distortion, intense ambient noise, and relatively low received power, a CDS cannot archive high data rate communication [1]. In order to overcome this drawback of CDS, relay communication techniques common in RFC systems can be considered in IROWC systems [16]-[17]. The signal from the transmitter can be enhanced by some relay terminals and forwarded to the user. Moreover, the relay terminals can enhance the received optical power at the user in situations where link blockage occurs between the transmitter and the user.

Figure 1 shows a typical two-phase relay mode used in the IROWC system. In this relay IROWC mode, a communication is divided into two phases. In the first phase, the infrared transmitter on the ceiling sends a signal to several relay terminals. In the second phase, the relay terminals send the signal to the users on the communication plane. However, for different relay terminals, the link distance from the transmitter via relay terminals to user are different causing dspersed signal arrival time at the

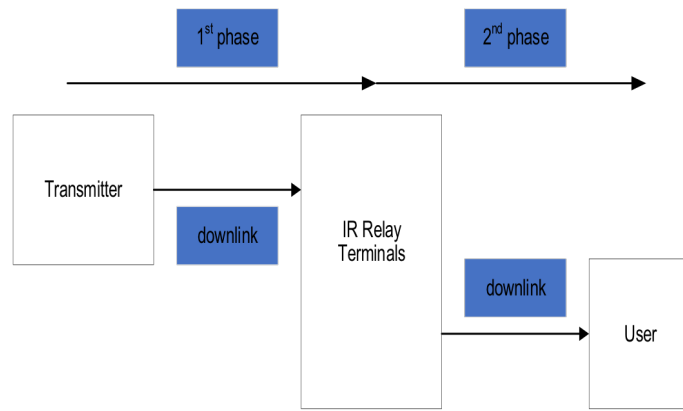

*Figure 1 the flow chart of relay communication in one time slot*

receiver. This multipath spread caused by multiple relay terminals can result in serious pulse spread and ISI.

In this paper, we propose a delay adaptation method to address this problem. This method balances the arrival time of the signals from different relay terminals by adding a pre-calculated (derived through a pre-distortion equalizer) delay to different relay terminals. In order to study the effectiveness of this method, we consider three different relay terminal deployment scenarios. The rest of the paper is organized as follows: The delay adaptation method is introduced in Section 2, while the simulation configuration and the results are provided in Section 3. Finally, the conclusions are presented in Section 4.

## 2. THE DELAY ADAPTATION METHOD

In order to overcome the drawback of different arrival time of signals in IROWC systems with relay terminals, a delay adaption method is used. This method reduces the time spread by adding differential delays to different beams, as first proposed in [18] - [24]. Before the formal communication begins, the relay terminals and the user synchronize at the start of a frame. Each relay terminal individually emits a special pulse sequence in a predetermined order. When the user receives the first pulse sequence, the user calculates the time delay time needed to achieve synchronization. The second relay terminal emits a pulse sequence after a predetermined time interval. The user receives this pulse sequence and hence determines the time delay of the second relay terminal. The differential delay is then determined. After the user computes the time delay of all relay terminals, a feedback signal of differential delays is sent to each relay terminal via an uplink channel.

An alternative method that can also be used to assist delay adaptation can be based on optical orthogonal codes. In the delay adaption method mentioned above, relay terminals and user synchronise at the start of a frame before the formal communication begin. However, each relay terminal of this new method simultaneously sends a special code based on a unique optical orthogonal code sequence. Users can identify the signals from different relay terminals with relevant decoders due to the auto and cross correlation properties of optical orthogonal codes [19]-[21].

The decoded signals are sent to a compartor circuit to compute the differential delay between relay terminals. After the user computes the time delay of all the relay terminals, a feedback signal that carries the differential delay information is sent to each relay terminal via an uplink channel. The delay adaption method based on optical orthogonal codes is shown in figure 2.

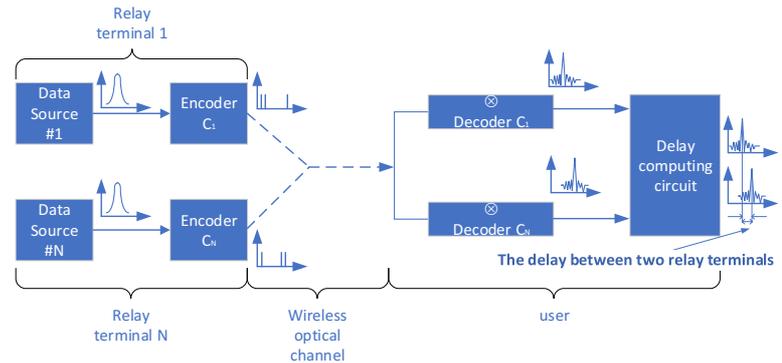

*Figure 2 delay adaption method based on optical orthogonal code technique*

## 3. SIMULATION SET-UP AND RESULTS

In order to evaluate the advantages of the proposed method, a simulation was performed using an empty room with dimensions of 8 m × 4 m × 3 m (length × width × height), see figure 3. The plaster walls are assumed to reflect light rays in a form close to a Lambertian function. Therefore, the walls (including the ceiling) and floor were modeled as Lambertian reflectors with reflectivity of 80 % and 30 %, respectively. Reflections from doors and windows are identical to reflections from walls. The transmitted signals are reflected from the room reflecting surfaces which were divided into a number of equal-size, square-shaped reflection elements. The reflection elements were treated as small transmitters that diffuse the received signals from their centres in the form of a Lambertian pattern [25]-[30]. It is noted that third-order and higher reflections do not produce a significant change in the received optical power, and therefore reflections up to second order are considered. The surface element size used in this work was set to 5 cm × 5 cm for the first-order reflections and 20 cm × 20 cm for the second-order reflections. The transmitter is located at the centre of the ceiling while the user is in the communication plane which is 1m above the floor. The room set-up is showed in figure 3.

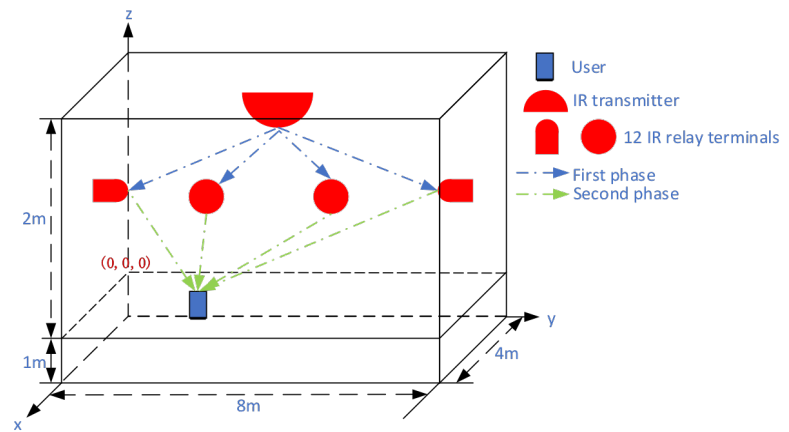

*Figure 3 room set up showing transmitter, receiver and relay nodes*

We consider three different scenarios for relay terminals deployment in which the relay terminals are separately deployed 0.5m, 1m and 1.5m under the ceiling around the walls. Each scenario contains 12 relay terminals. The mutual distance between them is 1 m. Additional simulation parameters are given in Table 1.

*Table 1. Simulation parameters*

| Parameter | Configuration |
|---|---|
| **Room set** | |
| Length, Width, Height | 8m, 4m, 3m |
| Reflectivity of walls | 0.8 |

| Reflectivity of ceiling | 0.8 | |
|---|---|---|
| Reflectivity of floor | 0.3 | |
| **IR Transmitter** | | |
| Quantity | 1 | |
| Location | (2,4,3) | |
| The average optical power | 1W | |
| hps | 60º | |
| **Relay terminals** | | |
| Quantity | 12 | |
| Locations | The relay terminals of three scenarios are separately deployed 0.5m, 1m and 1.5m under the ceiling around the wall. | |
| **Users** | | |
| Quantity | 1 | |
| Elevation | 90º | |
| Azimuth | 0º | |
| FOV | 90º | |
| **Resolution** | | |
| Time bin duration | 0.1ns | |
| Bounces | 1 | 2 |
| Surface elements | 32000 | 2000 |
| Wavelength | 850nm | |

In this simulation, the influence of relay terminal on intensity and waveform are ignored, the received signal for a user can be defined as follows:

$$I_r(t) = \sum_{r=n}^{N} R[X(t) \otimes h_{tr}(t) \otimes h_{ru}(t)], \quad (1)$$

where $N$ is the set of relay terminals, $R$ is the responsivity of the receiver, $X(t)$ is the instantaneous optical power at the transmitter, $h_{tr}(t)$ represents the channel impulse response for the channel between transmitter and relay terminal, $h_{ru}(t)$ represents the impulse response for the channel between relay and receiver, and $\otimes$ represents convolution. The root-mean-square delay spread (D), which is a good measure of signal pulse spread, can be given by [30]-[37]:

$$D = \sqrt{\frac{\sum (t_i - \mu)^2 Pr_i^2}{\sum Pr_i^2}}, \quad (2)$$

where $t_i$ is the delay time associated with the received optical power $Pr_i$, $\mu$ is the mean delay given by:

$$\mu = \frac{\sum t_i Pr_i^2}{\sum Pr_i^2} \quad (3)$$

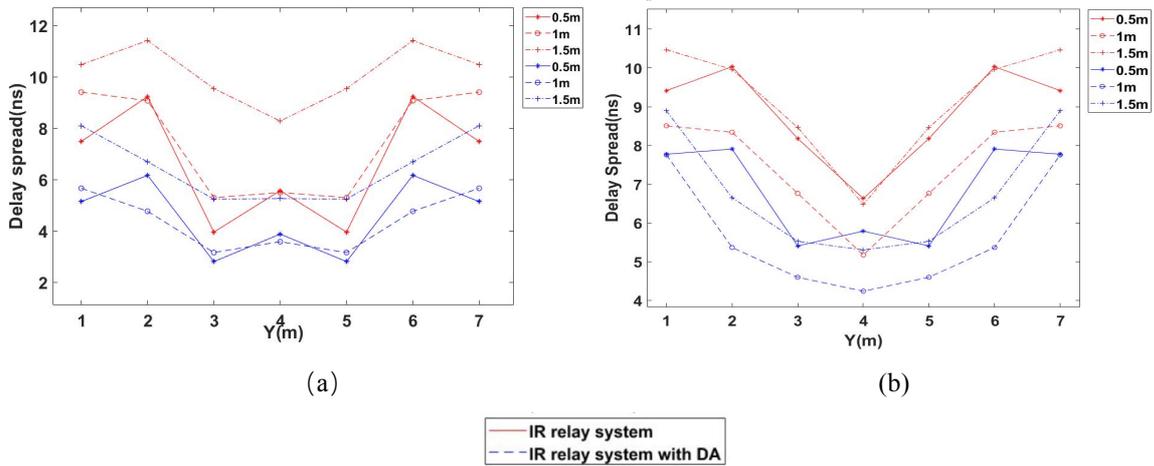

Figure 4 the root-mean-square delay spread results for (a) x=1m (b) x=2m

Figure 4(a) shows the root-mean-square delay spread (D) of conventional relay system and the relay system with the delay adaptation (DA) method when relay terminals are deployed at 0.5m 1m and 1.5m below the ceiling and user is moved along Y axis with x=1m, while Figure 4(b) shows the root-mean-square delay spread (D) results with x=2m.

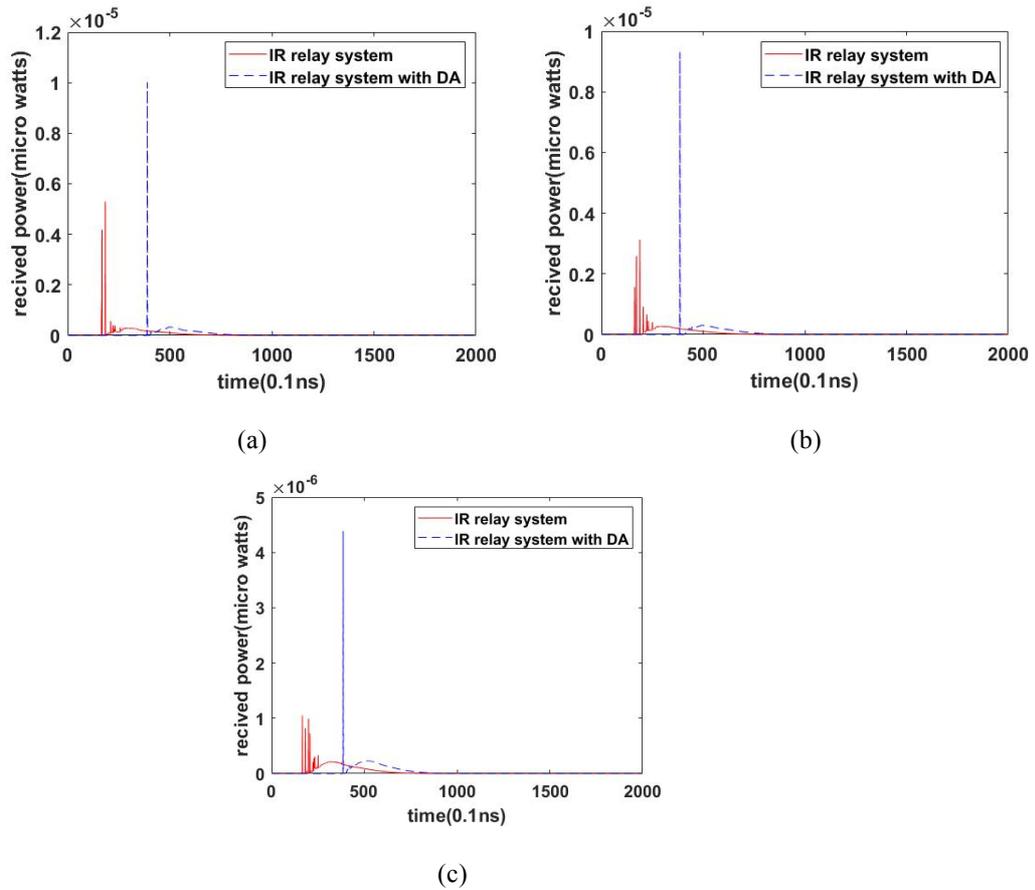

*Figure 5 the impulse response results in the conventional IR relay system and IR relay system with the delay adaptation (DA) method (a) Scenario 1 (b) Scenario 2 (c) Scenario 3*

Figure 5(a) shows the impulse response of the conventional relay system and relay system with delay adaptation method when the relay terminals are deployed at 0.5m and the user is at the room corner (1,1,1). Figure 5(b) shows the impulse responds of conventional relay system and relay system with the delay adaptation method when the relay terminals are deployed at 1 m and the user is at the room corner (1,1,1). Figure 5(c) shows the impulse responds of the conventional relay system and relay system with delay adaptation method when relay terminals are deployed at 1.5m and user at the room corner (1,1,1). The results show that the root-mean-square delay spread of the IR relay system with the DA method is lower than the conventional IR relay system in different scenarios. Meanwhile the pulse spread of the IR relay system is reduced by the DA method as the results showed in figure 5.

## 4. CONCLUSIONS

This paper introduced a delay adaptation method in IROWC relay systems. The root-mean-square delay spread and impulse response of conventional relay systems and relay systems with relay adaptation method are investigated in three different scenarios. The root-mean-square delay spread of relay systems with delay adaptation method is 30% less than that of the conventional relay system on average in three different scenarios. The relay system with delay adaptation method has better impulse response compared with conventional relay system. We have found that applying the delay adaptation method in IROWC relay systems reduces the delay spread. Future work includes analysing the SNR and optimizing relay terminal deployment schemes for the proposed system.

**ACKNOWLEDGMENTS**

The authors would like to acknowledge funding from the Engineering and Physical Sciences Research Council (EPSRC) INTERNET (EP/H040536/1), STAR (EP/K016873/1) and TOWS (EP/S016570/1) projects. SHM would like to thank EPSRC for providing her Doctoral Training Award scholarship. All data are provided in full in the results section of this paper.